\documentclass[acmsmall]{acmart}
\AtBeginDocument{%
  }

\setcopyright{acmlicensed}
\copyrightyear{2018}
\acmYear{2018}
\acmDOI{XXXXXXX.XXXXXXX}

\acmJournal{JACM}
\acmVolume{37}
\acmNumber{4}
\acmArticle{111}
\acmMonth{8}
\usepackage{multirow}
\usepackage{rotating}
\usepackage{hyperref}
\usepackage{algorithm}
\usepackage{algpseudocode}
\usepackage{bm}

\begin{document}

\title{JiuTian·Chuanliu: A Large Spatiotemporal Model for General-purpose Dynamic Urban Sensing}


\author{Liangzhe Han}
\affiliation{%
  \institution{CCSE, Beihang University}
  \city{Beijing}
  \country{China}}
\email{liangzhehan@buaa.edu.cn}

\author{Leilei Sun}
\affiliation{%
  \institution{CCSE, Beihang University}
  \city{Beijing}
  \country{China}}
\email{leileisun@buaa.edu.cn}

\author{Tongyu Zhu}
\affiliation{%
  \institution{CCSE, Beihang University}
  \city{Beijing}
  \country{China}}
\email{zhutongyu@buaa.edu.cn}


\author{Tao Tao}
\affiliation{%
  \institution{China Mobile Group IT Center}
  \city{Beijing}
  \country{China}}
\email{ttaoit@163.com}

\author{Jibin Wang}
\affiliation{%
  \institution{China Mobile Group IT Center}
  \city{Beijing}
  \country{China}}
\email{wangjibin@chinamobile.com}

\author{Weifeng Lv}
\affiliation{%
  \institution{CCSE, Beihang University}
  \city{Beijing}
  \country{China}}
\email{lwf@buaa.edu.cn}
\renewcommand{\shortauthors}{Han L et al.}

\begin{abstract}
As a window for urban sensing, human mobility contains rich spatiotemporal information that reflects both residents’ behavior preferences and the functions of urban areas. The analysis of human mobility has attracted the attention of many researchers. However, existing methods often address specific tasks from a particular perspective, leading to insufficient modeling of human mobility and limited applicability of the learned knowledge in various downstream applications. To address these challenges, this paper proposes to push massive amounts of human mobility data into a spatiotemporal model, discover latent semantics behind mobility behavior and support various urban sensing tasks. Specifically, a large-scale and widely covering human mobility data is collected through the ubiquitous base station system and a framework named General-purpose and Dynamic Human Mobility Embedding (GDHME) for urban sensing is introduced. The framework follows the self-supervised learning idea and contains two major stages. In stage 1, GDHME treats people and regions as nodes within a dynamic graph, unifying human mobility data as people-region-time interactions. An encoder operating in continuous-time dynamically computes evolving node representations, capturing dynamic states for both people and regions. Moreover, an autoregressive self-supervised task is specially designed to guide the learning of the general-purpose node embeddings. In stage 2, these representations are utilized to support various tasks. To evaluate the effectiveness of our GDHME framework, we further construct a multi-task urban sensing benchmark. Offline experiments demonstrate GDHME's ability to automatically learn valuable node features from vast amounts of data. Furthermore, our framework is used to deploy the JiuTian ChuanLiu Big Model, a system that has been presented at the 2023 China Mobile Worldwide Partner Conference and reported to use for smart tourism, urban planning and so on.
\end{abstract}



\received{-}

\maketitle

\section{Introduction}
In recent years, rapid urbanization and the swift expansion of metropolitan areas have dramatically reshaped modern cities. The large-scale mobility behavior of urban residents serves as the pulse of the city, encapsulating rich spatiotemporal information. This information not only reflects the behavioral preferences of residents but also provides valuable insights into the functional dynamics of urban areas. By analyzing these patterns, stakeholders can unlock critical opportunities to improve traffic management, optimize urban planning, and enhance public security. Furthermore, this data-driven approach has the potential to foster more sustainable and resilient urban environments, supporting long-term urban development and the well-being of city inhabitants.

\begin{figure}[!tbp]
    \centering
    \includegraphics[width=\textwidth]{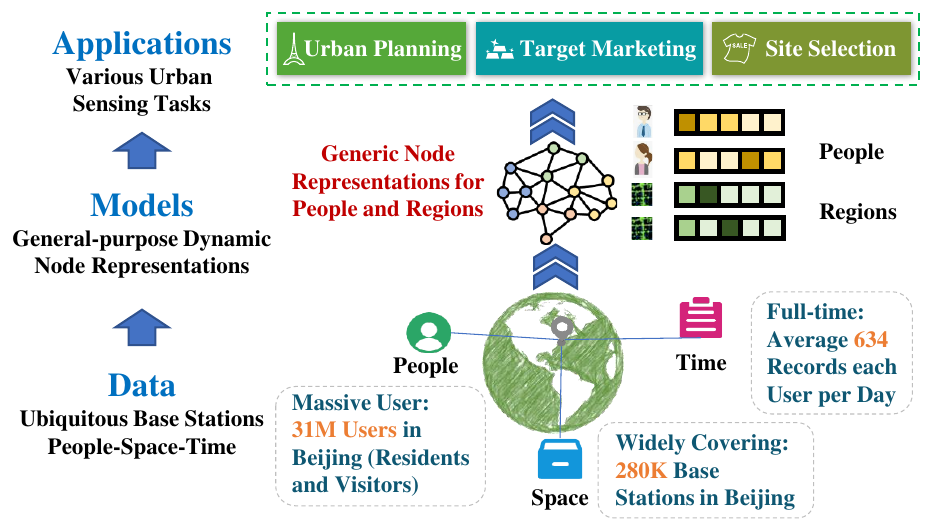}
    \caption{The overview of JiuTian·Chuanliu.}
    \label{fig: system}
\end{figure}

With the increasing availability of human mobility data and the advancements in deep learning, the analysis of human mobility has garnered significant attention from researchers.
Some researchers conceptualize human mobility as sequential trajectories of individuals, employing sequential and evolutionary learning methods to uncover spatiotemporal patterns embedded in these trajectories. Such approaches have been widely applied to tasks like next point-of-interest prediction and travel time estimation\cite{DBLP:conf/ijcai/Zhang0SZ18,DBLP:conf/aaai/WangZCLZ18,DBLP:conf/kdd/WangFY18,DBLP:conf/aaai/LinW0L21,DBLP:journals/tkde/ZhaoLLXLZSZ22}.
Other researchers focus on interpreting human mobility as semantic relationships between regions. By learning representation vectors for urban areas, these studies enable tasks such as land use analysis and annual crime rate prediction\cite{DBLP:conf/ijcai/WuYFPZZ0W22,wang2017region,DBLP:conf/ijcai/0004LLH20,DBLP:conf/www/ZhangHXWLY23,DBLP:conf/aaai/ZhouHCS023}. 
Additionally, a growing body of work extracts dynamic features of regions from human mobility data, leveraging spatiotemporal prediction techniques to forecast regional demand and origin-destination demand. These studies employ cutting-edge models, enabling precise forecasting for applications in urban planning and resource allocation\cite{yao2018deep, Deepresnet, revisiting, ye2019co,stgcn,dcrnn,gwnet,mtgnn,agcrn,astgcn}.
This diverse body of research highlights the potential of human mobility analysis in supporting a wide range of urban and societal applications, from enhancing transportation systems to improving public safety and resource management.

However, existing methods frequently focus on specific tasks from a singular perspective, which can result in an incomplete representation of human mobility and limit the transferability of the knowledge learned to other downstream applications. These task-specific approaches often fail to capture the multifaceted and interconnected nature of urban dynamics, reducing their effectiveness in addressing broader challenges.
In contrast to traditional paradigms, this paper introduces a novel approach that leverages self-supervised learning on large-scale human mobility data to enable general-purpose urban sensing. By adopting this method, the study seeks to construct versatile and comprehensive representations of human mobility, capable of supporting a wide range of urban applications.

Self-supervised learning has recently gained widespread attention and extensive application in fields like computer vision and natural language processing. Unlike traditional methods that rely on expert-crafted feature extraction or labor-intensive label collection, self-supervised learning constructs optimization tasks directly from the input data itself. By solving these tasks, models learn data representation vectors that encapsulate essential features of the original data, enabling a deeper and more comprehensive understanding of the underlying patterns.
These learned features have demonstrated remarkable versatility, allowing models to efficiently adapt to a wide range of downstream tasks. In computer vision, self-supervised learning has been employed to encode images, facilitating solutions to various image-related tasks such as image generation, recognition, and segmentation\cite{van2016conditional,razavi2019generating,chen2020simple,he2020momentum,grill2020bootstrap,he2022masked}. 
Similarly, in natural language processing, self-supervised learning on massive text corpora has empowered models to excel in diverse general-purpose tasks, including code generation, translation, and question answering\cite{DBLP:conf/nips/VaswaniSPUJGKP17,DBLP:conf/naacl/DevlinCLT19,radford2018improving,radford2019language,brown2020language,DBLP:conf/acl/ZhangWLB20,wei2022emergent}.This paradigm shift not only enhances task performance but also reduces dependency on labeled data, making self-supervised learning a cornerstone in the development of scalable, general-purpose AI systems across multiple domains.

Achieving self-supervised learning for urban sensing presents several significant challenges:
1. Identifying Key Data Representations: Determining the data elements that effectively capture human mobility remains an open question. Current approaches, which rely on trajectories or aggregated regional statistics, often lose critical information about the underlying mobility behaviors. This loss of detail can hinder the accurate representation of complex urban dynamics.
2. Modeling Spatiotemporal Dynamics: The continuous evolution of urban spatial regions and the people moving within them adds another layer of complexity. Capturing these dynamic interactions and their temporal variations is a major challenge, as traditional models often struggle to account for such fluid and multifaceted relationships.
3. Lack of Systematic Evaluation: Existing studies frequently focus on specific tasks, such as trajectory classification or land use analysis, without providing a comprehensive evaluation of the features learned for urban sensing. This task-centric approach limits our understanding of how effectively these features generalize across diverse urban applications.

To address the challenges of insufficient modeling and limited generalizability in existing mobility analysis, this work introduces a framework powered by the General-purpose and Dynamic Human Mobility Embedding (GDHME) method for urban sensing. 
The overview of the framework is shown in Figure \ref{fig: system}. 
GDHME collects large-scale human mobility data through the ubiquitous base station system. 
When a mobile device enters the coverage area of a base station, the communication is recorded, forming spatiotemporal trajectories of individuals. 
Owing to the wide coverage and continuity of such data, it provides a comprehensive foundation for general-purpose urban sensing.
After acquisition and preprocessing, the data is processed by the GDHME model in two stages. 
First, the framework encodes dynamic human mobility into evolving node representations that capture temporal states of people and regions. 
A self-supervised autoregressive task is designed to guide this representation learning without reliance on manual labels. 
Second, the learned embeddings are evaluated through a multi-task urban sensing benchmark and further deployed in real-world applications such as smart tourism and urban planning. 
The contributions of this paper can be summarized as following:\begin{itemize}
    \item \textbf{An urban sensing framework based on a large-scale spatiotemporal model is designed.} This framework defines human mobility as unified people-region-time interaction format and conduct self-supervised learning for urban sensing. It provides a private, efficient and effective paradigm bridging large-scale human mobility data and various urban sensing applications.
    \item \textbf{A bipartite representation learning method for people-region interaction data is proposed.} To the best of our knowledge, this is the first study to generate dynamic representations for both people and regions in this scale. With the sensing for these two basic entities, our method can support a wide range of downstream urban tasks.
    \item \textbf{A spatiotemporal platform named JiuTian·Chuanliu for urban sensing has been formed from the proposed framework.} The platform has been presented at the 2023 China Mobile Worldwide Partner Conference\footnote{\url{https://www.10086.cn/aboutus/news/groupnews/index_detail_47406.html}} and reported in a wide range of applications \footnote{\url{https://www.10086.cn/aboutus/news/groupnews/index_detail_49890.html}}\footnote{\url{https://new.qq.com/rain/a/20240525A060B200}}.
\end{itemize}

\section{Preliminaries}
\subsection{Human Mobility Dynamic Graph}
To extract fine-grained information from human mobility, we formalize the raw records as a continuous-time dynamic graph. 
Formally, the studied area is partitioned into a set of region nodes $\mathbb{V}^a=\{v_1^a,v_2^a,\cdots,v_N^a\}$, each corresponding to a non-overlapping region that collectively covers the whole city. 
Individuals in the dataset are represented as a set of person nodes $\mathbb{V}^p=\{v_1^p,v_2^p,\cdots,v_M^p\}$. 
When a person $v_i^p$ appears in region $v_j^a$ at time $t_k$ and stays for $\Delta t_k$, an interaction is recorded as a dynamic edge $e_k=(v_i^p,v_j^a,t_k,\Delta t_k)$. 
The dynamic graph at time $t$ is denoted as $\mathcal{G}(t)=(\mathbb{V}^a,\mathbb{V}^p,\mathbb{E}(t))$, where $\mathbb{E}(t)=\{e_k|t_k<t\}$ is the set of observed interactions up to $t$. 
The interaction sets related to a specific person or region are denoted as $\mathbb{E}^p_i(t)$ and $\mathbb{E}^a_j(t)$, respectively.

\subsection{Problem Definition}
Given the historical dynamic graph $\mathcal{G}(t)$ up to time $t$, our goal is to learn an encoder $f$ such that
\begin{equation}
    f(\mathcal{G}(t)) = (\bm{Z}^a, \bm{Z}^p),
\end{equation}
where $\bm{Z}^a \in \mathbb{R}^{N\times d}$ are the representations of all region nodes and $\bm{Z}^p \in \mathbb{R}^{M\times d}$ are the representations of all person nodes at time $t$. 
These embeddings are then utilized for downstream prediction tasks in stage 2:
\begin{equation}
    \hat{\bm{Y}}^a = g^a(\bm{Z}^a), \quad \hat{\bm{Y}}^p = g^p(\bm{Z}^p),
\end{equation}
where $g^a$ and $g^p$ are task-specific models for regions and people, respectively.

\section{Methodology}
\begin{figure*}[!htbp]
    \centering
    \includegraphics[width=\textwidth]{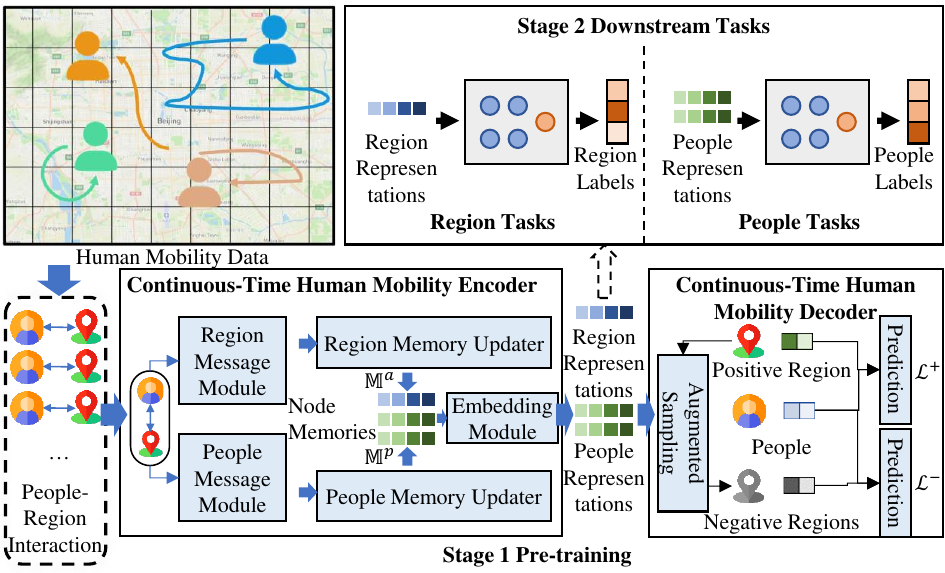}
    \caption{The overall pipeline of GDHME.
    The continuous-time human mobility encoder maintains memories for each region node and each people node.
    When people appear at new locations, the memories are updated according to interactions between people and regions.
    Then, the node representations are generated with the updated memories by the embedding module.
    Meanwhile, a continuous time human mobility decoder is specially designed to predict next location of the people in stage 1.
    In stage 2, the well-learned node representations will be utilized to solve various downstream tasks.}
    \label{fig: archtecture}
\end{figure*}
The overall pipeline of GDHME is illustrated in Figure \ref{fig: archtecture}, which can be roughly divided into two stages.
The aim of the first stage is to learn a general representation for each region node and each people node.
It is mainly accomplished with two modules: the continuous-time human mobility encoder and the continuous-time human mobility decoder.
\textbf{The encoder encodes the raw human mobility data into node representations to capture the information about regions and people behind human mobility.
The decoder takes the representation vectors as input to predict human mobility and optimize the representations with prediction loss.} 
And the aim of the second stage is to leverage the representations from the first stage and solve the various downstream tasks.
Specifically, to compress the information and dynamic states associated with one node into its node representations, the encoder maintains dynamic node memories for each node.
When a person appears at a certain location, the states of the person and the region are both affected with the newly-happened events.
Accordingly, the memories of these people and locations are updated with the events.
And the latest node states, i.e. dynamic node representations, can be read from these updated memories with the embedding modules.
To ensure that the node representations can contain meaningful information from the original data, a decoder to generate the future movement of the people is employed.
And after pre-training in the first stage, the model can generate dynamic representations.
When we want to solve region-related tasks, the node representations for regions can be fetched and inputted to a relatively simple model to reach the target.
And the procedure for people-related tasks and people-region-interacted tasks are similar.
\subsection{Continuous-time Human Mobility Encoder}
In the real world, the states of the city are evolving all the time.
For example, people always go back to their houses at night for sleeping and go to work in the daytime.
As for the regions, the situation where many people tend to go to work in the morning may cause morning peak in living areas, and the shopping areas may be busier in the weekends than weekdays.
The states of city may be complex and hard to analysis at the first glance.
However, all elements behind the city are associated with the people and regions, and their association can be revealed in the human mobility.
Thus, how to deal with human mobility data is the key to sense the states of the city.

In existing studies, researchers always process the data as certain formats.
Some researchers extract region features with human mobility data, such as inflow, outflow.
This process would drop fine-grained information from the original data and lead to limited application.
Some researchers focus on the people view and model individual trajectories as location sequences.
These methods always fail to model the dynamics and abundant semantics for locations.
In our opinion, all these processing are based on the same data and these data is generated from human moving between different locations.
Thus, the basis of modeling region and people can be unified as one type of data and the basic unit of this type of data can be viewed as human-region interaction.
In other words, when a person arrives at a region, it generate an interaction which can affect both of them.
In people view, their states are the combination of their historical visited locations and in region view, their states are the combination of people who visit them.
For example, when a person arrive a area and stay there for 8 hours at night, the states of the person may be sleeping here.
And if many people gather at a place in the weekdays, it may be a school or a working area.

\begin{figure}
    \centering
    \includegraphics[width=\columnwidth]{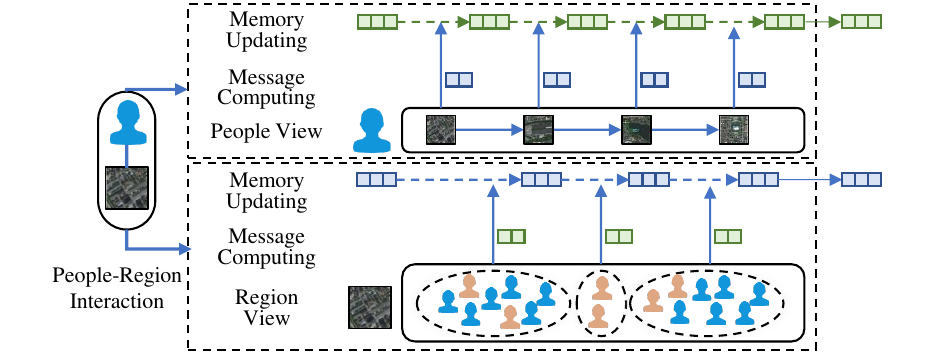}
    \caption{Illustration of the continuous-time human mobility encoder.}
    \label{fig: encoder}
\end{figure}

To achieve this goal, a continuous-time human mobility encoder is proposed here to learn dynamic node representation from the original people-region interaction data.
The fundamental idea is to maintain memories as node states and update them incrementally according to the newly happened events.
The overall procedure is shown in Figure \ref{fig: encoder}.
The procedure is achieved with the message computing modules, the memory updating modules and the node embedding modules.
First, when interactions happen, the messages for relative people and regions are computed:
\begin{equation}
    \bm{m}^p_i(t_k),\bm{m}^{a}_j(t_k) = MESS(\mathbb{M}^p_i(t^{p-}_i),\mathbb{M}^a_j(t^{a-}_j),e_k),
\end{equation}
where $e_k=(v_i^p, v_j^a, t_k)$ is the interaction happen between people $v_i^p$ and region $v_j^a$ at time $t_k$, $\bm{m}^p_i(t_k)$ is the message for people $v_i^p$ at time $t_k$, $\bm{m}^a(t_k)$ is the message for region $v_j^a$ at time $t_k$, $\mathbb{M}^p_i(t^{p-}_i)$ is the memories for people $v_i^p$ at last update time $t^{p-}_i$ and $\mathbb{M}^a_j(t^{a-}_j)$ is memories for region $v_j^a$ of last update.
Then these messages are used to update corresponding memories as follows:
\begin{equation}
    \mathbb{M}^p_i(t_k),\mathbb{M}^a_j(t_k) = UPD(\bm{m}^p_i(t_k),\bm{m}^{a}_j(t_k), \mathbb{M}^p_i(t^{p-}_i),\mathbb{M}^a_j(t^{a-}_j)),
\end{equation}
where $\mathbb{M}^p_i(t_k)$ are the updated memories for people $v_i^p$ and $\mathbb{M}^a_j(t_k)$ are the updated memories for region $v_j^a$.
And the node representations are obtained from these latest memories:
\begin{equation}
    \bm{Z}^p(t),\bm{Z}^a(t) = EMB(\mathbb{M}^p_i(t),\mathbb{M}^a_j(t)).
\end{equation}

Compared to transformer-based models, which typically treat the sequence of visited regions as static tokens and fail to capture the evolving semantics of regions influenced by diverse human interactions, our bipartite encoder dynamically updates both people and region states at each interaction.
Compared to spatiotemporal models, which primarily focus on aggregated region-level flows and cannot reflect individual-specific semantic signals, our approach explicitly integrates user- and region-level dynamics in a unified framework.
This dual updating mechanism allows our model to better capture the intertwined evolution of people and urban spaces, thereby providing more general-purpose and adaptive embeddings for downstream urban sensing tasks.
\subsubsection{Node Memory}
As stated above, the role of the node memories is to compress the historical interaction into vectors.
There are three basic guidelines: first, the current states should result from the historical interaction information; second, for dynamic node representations, the more recent an interaction is, the more influence it will have on current node representations; third the longer an interaction is, the more important it is for node representations.
Meanwhile, there are some attributes of people and regions remain the same for different time.
Thus, the people memories are designed with two parts: static memories and dynamic memories:
\begin{equation}
    \mathbb{M}^p_i(t)=(\bm{M}_i^{p,d}(t),\bm{M}_i^{p,s}(t),t^{p-}_i),
\end{equation}
where $t^{p-}_i$ is the last update time of $v^p_i$.
For the dynamic memories for people $v^p_i$ at time $t$, it is computed as the time-decaying weighted combination of historical messages for $v^p_i$:
\begin{equation}
  \bm{M}_i^{p,d}(t)=\frac{\sum\limits_{e_k\in \mathbb{E}^p_i(t)}exp(- (t-t_k))\bm{m}^p_i(t_k)}{\sum\limits_{e_k\in \mathbb{E}^p_i(t)}exp(-(t-t_k))}.
  \label{people_dynamic_rep}
\end{equation}
According to this equation, the memories will remember the historical interactions and assign more weights on the recent events to learn the dynamic states of people.
For the static memories, it is computed as average combination of historical messages for $v^p_i$:
\begin{equation}
  \bm{M}_i^{p,s}(t)=\frac{\sum\limits_{e_k\in \mathbb{E}^p_i(t)}\Delta t_k\bm{m}^p_i(t_k)}{\sum\limits_{e_k\in \mathbb{E}^p_i(t)}\Delta t_k}.
  \label{people_static_rep}
\end{equation}
The intuition of the static representations is to learn the relatively static patterns from multiple days activities.
For example, if a person spent most time at two points, this person may be a student or a worker, and if a person spent most time hanging around, this person tend to be a traveller or taxi driver.

As for region nodes, they also have dynamic states and static attributes.
For example, the information that an area is business area is static while how many people will come to a business area and the house price is dynamic.
Similarly, the memories of region node $v^a_j$ also contain the static parts and the dynamic parts:
\begin{equation}
  \bm{M}_j^{a,d}(t)=\frac{\sum\limits_{e_k \in \mathbb{E}^a_j(t)}exp(- (t-t_k))\bm{m}^a_j(t_k)}{\sum\limits_{e_k \in \mathbb{E}^a_j(t)}exp(-(t-t_k))},
  \label{region_dynamic_rep}
\end{equation}
\begin{equation}
  \bm{M}_j^{a,s}(t)=\frac{\sum\limits_{e_k \in \mathbb{E}^a_j(t)}\Delta t_k\bm{m}^a_j(t_k)}{\sum\limits_{e_k \in \mathbb{E}^a_j(t)}\Delta t_k}.
  \label{region_static_rep}
\end{equation}

Directly computing memories above will lead to a time-consuming procedure.
Here, we decompose these memories as two parts as:
\begin{equation}
  \bm{M} = \frac{\bm{A}}{B},
\end{equation}
where $\bm{A}$ is the numerator part and $B$ is the denominator part.
And they can then be updated incrementally with the following message computing module and memory updating module.
\subsubsection{Message Computing Module}
For region nodes, their states is determined by the amount and the type of people come here.
Thus, the message for region nodes is computed with the above dynamic and static memories of people.
Formally, when there is an interaction $e_k=(v_i^p, v_j^a, t_k, \Delta t_k)$ the messages are computed as:
\begin{equation}
  \bm{m}^a_j(t_k)=MESS\_FUN(\bm{M}_i^{p,d}(t_k),\bm{M}_i^{p,s}(t_k)),
\end{equation}
where $MESS\_FUN$ is the message functions, such as mean pooling, Multi-Layer Perceptron.

For people node, its states is where the people travel before, and the message to update it is computed with the above dynamic and static memories of regions.
And the messages for node $v_i^p$ are computed as follows:
\begin{equation}
  \bm{m}^p_i(t_k)=MESS\_FUN(\bm{M}_j^{a,d}(t_k),\bm{M}_j^{a,s}(t_k), \bm{E}_j),
\end{equation}
where $\bm{E}_j\in \mathbb{R}^d$ is parameterized embeddings for region node $v^a_j$.

\subsubsection{Memory Updating Module}
After the messages for certain nodes are collected, the memories of these nodes are updated.
The basic idea is to combine the existing memories from the last update and messages from newly happened events.
This incremental procedure can be achieved by separately updating the $\bm{A}$ and $B$.
Formally, after omitting the symbol of people and region, the updating procedure for dynamic memories is as follows:
\begin{equation}
    \begin{aligned}
  \bm{A}_i^d(t_k)&=exp(- (t_k-t_i^-))\bm{A}_i^d(t_i^-) + \bm{m}_i(t_k),\\
    B_i^d(t_k)&=exp(- (t_k-t_i^-))B_i^d(t_i^-) + 1,
  \end{aligned}
\end{equation}
where the decaying item before last updated $\bm{A}$ assures that dynamic memories can assign more weights on recent events to sense dynamic node states.
The static memories are updated as:
\begin{equation}
    \begin{aligned}
  \bm{A}_i^s(t_k)&=\bm{A}_i^s(t_i^-) + \Delta t_k\bm{m}_i(t_k),\\
    B_i^s(t_k)&=B_i^s(t_i^-) + \Delta t_k,
  \end{aligned}
\end{equation}
where no decaying item in static memories updating make the static memories can focus on static and long-term attributes of nodes.
And the update time of people and regions is also updated:
\begin{equation}
    t_i^-\leftarrow t_k.
\end{equation}

In summary, these two updating mechanism provide temporal-local and temporal-global view for nodes to obtain more comprehensive node representations.

\subsubsection{Embedding Module}
After update the memories to the latest ones, the dynamic node representations can be obtained from them.
First, the initial node representation are read from these memories:
\begin{equation}
    \begin{aligned}
    \bm{Z}^{p^\prime}_i(t)=MEAN(\bm{M}_i^{p,d}(t),\bm{M}_i^{p,s}(t)),\\
    \bm{Z}^{a^\prime}_j(t)=MEAN(\bm{M}_j^{a,d}(t),\bm{M}_j^{a,s}(t), \bm{E}_j).
  \end{aligned}
\end{equation}
The initial embeddings contains node information to some extent, but it is easy to be disturbed.
For example, the dynamic people node representations assign more weights on recently happened events.
However, due to the device problem, the event may be unreliable.
Thus, we further aggregate information from node historical neighbors for more robustness.
The aggregation is achieved adaptively, and the final people representation is computed as:
\begin{equation}
    \begin{aligned}
    \bm{Z}^{p}_i(t)=\sum\limits_{e_k \in \mathbb{E}^a_j(t)^\prime}\alpha_k Linear(\bm{Z}^{a^\prime}_j(t)),\\
   \alpha_k=\frac{Linear(\bm{Z}^{a^\prime}_j(t))Linear(\bm{Z}^{a^\prime}_j(t))}{\sum\limits_{e_k \in \mathbb{E}^a_j(t)^\prime}Linear(\bm{Z}^{a^\prime}_j(t))Linear(\bm{Z}^{a^\prime}_j(t))^T},
  \end{aligned}
\end{equation}
where $\mathbb{E}(t)^\prime$ is the sampling set of corresponding nodes.
This procedure aggregates multiple historical region node into people node representations for a better understanding of people states.
Similarly, the region node representations also aggregate multiple people nodes for a robust embedding.

\subsection{Continuous-time Human Mobility Decoder}
\begin{figure}
    \centering
    \includegraphics[width=\columnwidth]{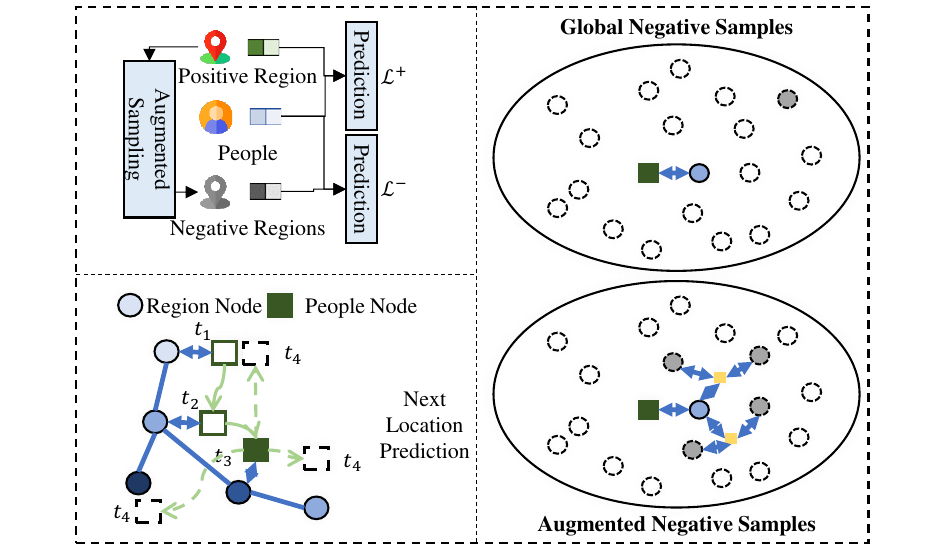}
    \caption{Illustration of the continuous-time human mobility decoder.}
    \label{fig: decoder}
\end{figure}
In the above section, a continuous-time human mobility encoder is designed to compress interaction information into node representation vectors.
To learn meaningful representations for people and regions, another crucial part is the optimization target.
The target should be task-agnostic and reflect the latent patterns in data. And the labels should be easy to obtain.
In this paper, a widely applied idea, generating the data itself, is adopted in the first stage.
In other words, the task is to predict which region the people visit next.

The generative task can effectively learn meaningful node representations, because it rely on a precise identification of node states.
For example, if a person sleeps at home, the next location should still be the same place as before, while if a person is on the way to work, the next location should be the region on the destination side.
Similar for the region states, if a region node contains many restaurants, it may be more likely to attract people in the lunch time.
However, designing a decoder for next location prediction is non-trivial.
First, the amount of candidate region nodes is large.
If the possibility score is computed between people node with all updated region nodes, the time-consuming is very expensive.
Second, the value of a very large part of region nodes is unhelpful as the negative nodes.
As shown in Figure \ref{fig: decoder}, one node is very distant from last travelled node, it is less possible for a person to travel it.
These negative nodes can be easily judged by the distance and can only bring limited supervision signals to learn.

In this paper, a continuous-time human mobility decoder is specifically designed to address these issues.
Basically, the task is to predict which location a person will travel next.
To alleviate the computing burden, instead of predicting the person with the whole set of region nodes, the idea of negative sampling is adopted.
However, if the negative samples are randomly sampled from all regions, it will be so easy to distinguish by distance.
To avoid the trivial solutions caused by naive negative samples, a harder sampler is designed.
Specifically, the sampled region node for harder negatives is sampled from two-hop neighbors of the positive region node.
If the model can distinguish the harder negatives from the positive, it need to learn more semantic information beyond the naive distance relation.

Formally, for a future event $e_{k}^+=(v_i^p, v_j^a, t_k^+,\Delta t_k^+)$, the negative sample is denoted as $e_{k}^{-}=(v_i^p, v_n^{a}, t_k^+,\Delta t_k^+)$.
By default, the negative node is sampled from the whole set of region nodes.
Meanwhile, the negative node is also sampled from the two-hop neighborhood with $p^n$ possibility:
\begin{equation}
    v_n^{a}\in\left\{ 
    \begin{array}{cc}
         \mathbb{V}^a,& 1-p^n \\
         \{v^a_i|(v^p_k, v^a_j,\cdot,\cdot) \in \mathbb{E}\wedge(v^p_k, v^a_i,\cdot,\cdot) \in \mathbb{E}\},& p^n
    \end{array}
    \right.
\end{equation}
The possibility for the future event $e_{k}^+$ is computed as:
\begin{equation}
    p(e_{k}^{+})=Sigmoid(MLP(\bm{Z}^p_i + \bm{Z}^a_j)).
\end{equation}
And the loss function is computed as:
\begin{equation}
    \mathcal{L}=-log(p(e_{k}^+))-log(1-p(e_{k}^{-})).
\end{equation}
\subsection{Downstream Tasks}
After the pre-training in the first stage, the continuous-time people and region representations can be generated based on the human mobility data.
This procedure can extract useful and latent patterns from the raw data.
And training a relatively simple model on these node representations can solve the downstream tasks.
For the region-centric tasks such as the region usage prediction, the model can be established on the latest region node representations:
\begin{equation}
    \hat{\bm{Y}^a}=MLP(\bm{Z}^a),
\end{equation}
where the $\hat{\bm{Y}^a}$ is the prediction of downstream tasks.
The procedure is similar for the people-centric tasks.

The first advantage of this pipeline is that it is more label-efficient, which means it can reach a satisfactory performance with relatively less labels.
This characteristic is very useful in the situation where the label is expensive.
The second advantage of this pipeline is that it can empower various applications with less efforts.
As the pre-training finish the work to learn basic features from human mobility data, the downstream tasks can be easily solved with the well-learnt node representations.
The third advantage of this pipeline is that it can protect privacy.
Without releasing the raw human mobility data, the data owner can release node representations for partners.
\section{Experiments}
This section will introduce experiments conducted to evaluate GDHME. The source code will be available upon acceptance.
\begin{table}[!htbp]
\centering
\caption{Statistic information of the datasets.}
    \begin{tabular}{cccc}
    \hline
    \# People & \# Regions & \# Interaction & Time Span \\
    \hline
    31,481,914 & 34,445 & 19,983,810,486 & 2023.3.1-2023.3.10 \\
    \hline
    \end{tabular}
    \label{table: datasets}
\end{table}
\subsection{Dataset}
The experiments are conducted on a large-scale signaling data collected in Beijing by China Mobile.
The dataset is generated when the cell phones communicate with the base station.
The communication record can reflect when a person enter and leave a area.
The dataset covers all the base stations of the city and all the time from 2023.3.1 to 2023.3.10.
The whole city is divided into $0.002^{\circ}\times0.002^{\circ}$ grids by the latitude and the longitude.
Each grid is viewed as a region node and each person is viewed as a people node.
More details are listed in Table \ref{table: datasets}.

\begin{figure}[!htbp]
    \centering
    \includegraphics[width=\columnwidth]{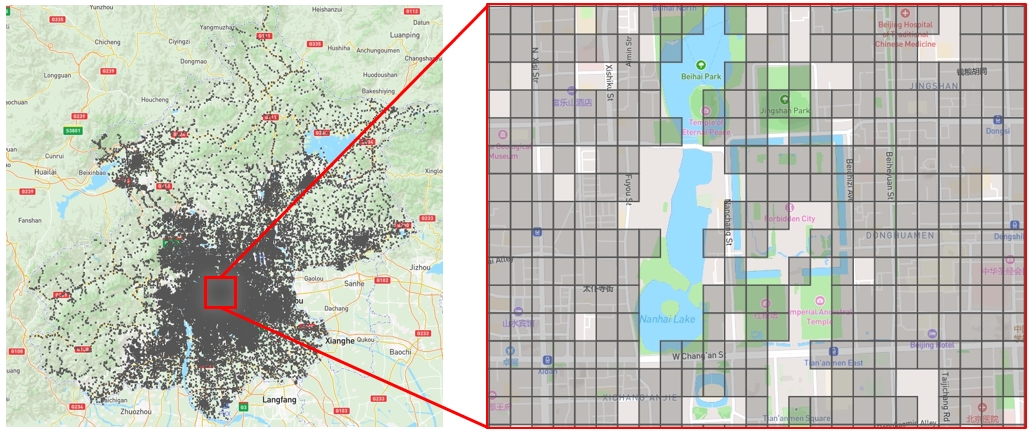}
    \caption{Illustration of the regions.}
    \label{fig: gridding}
\end{figure}
We first map all base stations into city-wide grids and remove grids without base stations. The final grid count is 34,445, with denser coverage in downtown areas. This gridding reduces the training burden by decreasing the number of region nodes and the associated memory and parameterized features. Additionally, gridding mitigates noise arising from base station traffic load balancing. To further clean the data, multiple signaling records within the same grid are merged, aggregating dwelling times. Spatial drift anomalies are filtered by examining speed between consecutive points: any intermediate point where the speed between three successive locations exceeds 144 km/h is treated as anomalous and removed. The resulting data serve as input to the GDHME model.
\subsection{Experiment Setups}
We implement GDHME with Python (3.8.11) and Pytorch (1.12.1) on a machine with 8 Nvidia A100 GPUs. 
In stage 1 (pre-training): 
Adam optimizer and the maximum training epochs is 20. 
The learning rate is 0.0001. 
The memory dimension and message dimension are 512. 
The possibility of harder negative sampling $p^n$ is set as 70\%.
The best parameters on the validation set are chosen to generate the node representations on the whole dataset.

In stage 2 (downstream tasks),
Mean Average Error (MAE), Root Mean Square Error (RMSE), Pearson Correlation Coefficient (PCC) are as metrics for regression tasks. The lower MAE and RMSE are better, and the higher PCC is better. AUC is the metric for binary classification tasks. F1-score is the metric for multi-class classification tasks. The higher AUC and F1-score is better.
\subsection{A Multi-task Benchmark}
To comprehensively evaluate the performance of representations learnt from human mobility, we introduce a multi-task benchmark, a collection of tasks across a diverse set about people and regions.
The aim of the proposed benchmark is to evaluate how much information can be extracted from the massive location data. Thus, the benchmark mainly covers the tasks about people and location profiling. A good performance on the benchmark may related to several real-world applications, such as precise marketing, region status forecasting and region risk detection.
The descriptions and statistics of these tasks are shown in Table \ref{table: tasks}.
\begin{table}[!htbp]
\centering
\caption{Task descriptions and statistics.}
  \begin{tabular}{c|ccc}
    \hline
   Type & Task & \# Samples & Target\\
   \hline
   \multirow{9}{*}{\rotatebox{90}{Region Function}} & Housing Price & 6,159 & Regression\\
    & \#Shopping PoI & 34,445 & Regression\\
    & \#Living PoI & 34,445 & Regression\\
    & \#Traffic PoI & 34,445 & Regression\\
    & \#Food PoI & 34,445 & Regression\\
    & \#GOV PoI & 34,445 & Regression\\
    & \#Company PoI & 34,445 & Regression\\
    & \#Car PoI & 34,445 & Regression\\
    & MostPoI & 34,445 & 23-class Classification\\
    \hline
    \multirow{6}{*}{\rotatebox{90}{People Profiling}}
    & Commuter & 47,498 & Binary Classification\\
    & Ride Hailing Driver & 2,706 & Binary Classification\\
    & House Owner & 10,000 & Binary Classification\\
    & Vehicle Owner & 10,000 & Binary Classification\\
    & APP Preference & 10,000 & Binary Classification\\
    & Traffic Preference & 10,000 & 5-class Classification\\
   \hline
  \end{tabular}
    \label{table: tasks}
\end{table}
\section{Downstream Task Comparison}
\begin{table*}[ht!]
    \caption{Performance of GDHME on region-centric tasks.}
    \centering
    \resizebox{\columnwidth}{!}{
    \begin{tabular}{c|c|c|cccc|ccc}
    \toprule
    Task & Stage 2 & Metric & Random & Feature & Node2Vec & TGN & w EasyDec &  w/o Static & GDHME\\
    \midrule
    \multirow{6}{*}{\rotatebox{90}{\textbf{Housing Price}}} & \multirow{3}{*}{LR}& MAE &34569.2949 &30167.6879 &18545.7781  &19025.5831 &17856.5357 &23723.9464 & \textbf{16645.5902 } \\
    && RMSE &41603.0047 &36969.9727 &25722.6132 &25226.0217 &24724.2623 &30651.1964 & \textbf{22970.0699 } \\
    && PCC &0.0373 &0.2527 &0.7526 &0.7340 &0.7506 &0.5984 & \textbf{0.7847 } \\
    \cline{2-10}
    & \multirow{3}{*}{MLP} & MAE &33857.4574 &55936.7088 &17985.1722 &18492.0943 &16968.8711 &25654.7407 & \textbf{15395.8587 } \\
    && RMSE &40987.6551 &69045.1750 &25538.3795 &26233.3614 &24614.3610 &32876.4472 & \textbf{22401.9195 } \\
    && PCC &0.0243 &0.0829 &0.7295 &0.7402 &0.7725 &0.5150 & \textbf{0.8072 } \\
    \hline
    \multirow{6}{*}{\rotatebox{90}{\textbf{\#Shopping PoI}}} & \multirow{3}{*}{LR} & MAE &9.3534 &8.8845 &8.6456 &8.5403 &8.2480 &8.7782 & \textbf{8.1210 } \\
    && RMSE &12.7699 &12.2313 &11.9189 &11.8148 &11.4138 &12.0594 & \textbf{11.4047 } \\
    && PCC &0.0069 &0.1330 &0.2613 &0.3415 &0.3563 &0.1995 & \textbf{0.4178 } \\
    \cline{2-10}
    & \multirow{3}{*}{MLP} & MAE &10.5752 &8.7742 &8.9155 &7.3793 &6.2963 &9.5848 & \textbf{6.2724 } \\
    && RMSE &14.2528 &12.4081 &12.5183 &10.8853 &9.4256 &13.0445 & \textbf{9.2180 }\\
    && PCC &0.0118 &0.1168 &0.4164 &0.5387 &0.6776 &0.2187 & \textbf{0.6870 } \\
    \hline
    \multirow{6}{*}{\rotatebox{90}{\textbf{\#Living PoI}}} & \multirow{3}{*}{LR} & MAE &7.4523 &7.3148 &6.9097 &6.3294 &6.4904 &6.8134 & \textbf{6.2324 }\\
    && RMSE &5.9628 &5.8081 &5.3820 &4.8893 &5.0530 &5.3385 & \textbf{4.7834 }\\ 
    && PCC &0.0096 &0.2074 &0.4175 &0.5101 &0.4966 &0.3855 & \textbf{0.5461 }\\  
    \cline{2-10}
    & \multirow{3}{*}{MLP}&MAE &8.6760 &7.5858 &7.0395 &5.9244 &5.1600 &7.3893 & \textbf{5.1198 } \\
    &&RMSE &6.8867 &5.9325 &5.3239 &4.3702 &3.6785 &5.8177 & \textbf{3.6689 } \\
    &&PCC &0.0005 &0.1695 &0.4874 &0.6268 &0.7272 &0.3395 & \textbf{0.7387 } \\
    \hline
    \multirow{6}{*}{\rotatebox{90}{\textbf{\#Traffic PoI}}} & \multirow{3}{*}{LR}& MAE &6.9443 &6.5680 &6.3193 &5.9670 &6.0756 &6.4570 & \textbf{5.8733 } \\
    && RMSE &5.5478 &5.1739 &4.8619 &4.5531 &4.6627 &4.9898 & \textbf{4.4938 } \\
    && PCC &0.0104 &0.2137 &0.4226 &0.4881 &0.4794 &0.3748 & \textbf{0.5216 } \\
    \cline{2-10}
    & \multirow{3}{*}{MLP} &MAE &8.1632 &6.8651 &6.6510 &5.4396 &5.0207 &7.0120 & \textbf{4.8471 } \\
    &&RMSE &6.4245 &5.2599 &4.9973 &4.0273 &3.6414 &5.3738 & \textbf{3.5559 } \\
    &&PCC &0.0049 &0.2692 &0.4757 &0.6330 &0.7037 &0.3225 & \textbf{0.7251 } \\
    \hline
    \multirow{6}{*}{\rotatebox{90}{\textbf{\#Food PoI}}} & \multirow{3}{*}{LR} &MAE &7.2825 &7.3545 &6.8808 &6.8656 &6.8093 &6.9633 & \textbf{6.7018 } \\
    &&RMSE &5.5738 &5.6226 &5.1762 &5.1289 &5.1239 &5.2893 & \textbf{5.0964 } \\
    &&PCC &0.0049 &0.1292 &0.2866 &0.3482 &0.3563 &0.2399 & \textbf{0.3643 } \\
    \cline{2-10}
    & \multirow{3}{*}{MLP} &MAE &8.4358 &17.2023 &7.7709 &6.8100 &5.8445 &7.6177 & \textbf{5.7351 } \\
    &&RMSE &6.4707 &11.0624 &5.6860 &4.7965 &4.0937 &5.6970 & \textbf{3.9573 } \\
    &&PCC &0.0107 &0.1059 &0.3349 &0.4632 &0.6365 &0.2013 & \textbf{0.6467 } \\
    \hline
    \multirow{6}{*}{\rotatebox{90}{\textbf{\#GOV PoI}}} & \multirow{3}{*}{LR} & MAE &6.8324 &6.6475 &6.3162 &6.5073 &6.4939 &6.6264 & \textbf{6.1548 } \\
    && RMSE &5.2932 &5.1507 &4.9009 &5.0630 &5.0461 &5.1294 & \textbf{4.7271 } \\
    && PCC &0.0019 &0.0859 &0.2225 &0.2446 &0.2525 &0.1037 & \textbf{0.3099 } \\
    \cline{2-10}
    & \multirow{3}{*}{MLP} &MAE &7.5564 &6.8867 &7.3216 &6.3753 &5.7018 &7.1978 & \textbf{5.4874 }\\
    &&RMSE &5.7796 &5.2368 &5.5313 &4.7422 &4.1819 &5.5350 & \textbf{3.9589 } \\
    &&PCC &0.0066 &0.0348 &0.2999 &0.4162 &0.5735 &0.1107 & \textbf{0.6034 } \\
    \hline
    \multirow{6}{*}{\rotatebox{90}{\textbf{\#Company PoI}}} & \multirow{3}{*}{LR}& MAE &7.8629 &7.9955 &7.4738 &7.4546 &7.4901 &7.7861 & \textbf{7.2804 } \\
    && RMSE &5.6965 &5.8115 &5.3181 &5.3904 &5.4396 &5.6381 & \textbf{5.1429 } \\
    && PCC &0.0180 &0.1152 &0.2957 &0.3124 &0.3301 &0.1456 & \textbf{0.4130 } \\
    \cline{2-10}
    & \multirow{3}{*}{MLP}&MAE &8.7705 &8.4287 &7.4293 &6.3909 &5.5724 &8.1047 & \textbf{5.3363 } \\
    &&RMSE &6.3657 &6.2333 &5.1244 &4.2043 &3.5466 &5.7967 & \textbf{3.4862 } \\
    &&PCC &0.0124 &0.0392 &0.4572 &0.6157 &0.7243 &0.1937 & \textbf{0.7468 } \\
    \hline
    \multirow{6}{*}{\rotatebox{90}{\textbf{\#Car PoI}}} & \multirow{3}{*}{LR}& MAE &26.5266 &25.2450 &22.5612 &19.7490 &20.0524 &22.3556 & \textbf{17.7758 } \\
    && RMSE &18.8792 &18.5106 &15.6648 &14.1609 &14.0838 &16.1997 & \textbf{11.4634 } \\
    && PCC &0.0017 &0.2387 &0.4983 &0.6503 &0.6344 &0.5051 & \textbf{0.7342 } \\
    \cline{2-10}
    & \multirow{3}{*}{MLP}&MAE &30.0339 &24.6313 &20.6714 &17.6526 &14.9527 &22.6332 & \textbf{14.7770 } \\
    &&RMSE &18.4348 &16.3871 &10.7744 &8.7622 &6.6600 &12.3072 & \textbf{6.5359 } \\
    &&PCC &0.0104 &0.3666 &0.6310 &0.7463 &0.8295 &0.5365 & \textbf{0.8308 } \\
    \hline
    \multirow{4}{*}{\rotatebox{90}{\textbf{MostPoI}}} & \multirow{2}{*}{KNN}& micro-F1 &0.1692 &0.2498 &0.4568 &0.4699 &0.5594 &0.1869 & \textbf{0.5676 } \\
    && macro-F1 &0.0555 &0.0989 &0.2937 &0.3061 &0.4071 &0.0190 & \textbf{0.4185 } \\
    \cline{2-10}
    & \multirow{2}{*}{MLP}& micro-F1 &0.1402 &0.3124 &0.3600 &0.4484 &0.5180 &0.2550 & \textbf{0.5421 } \\
    && macro-F1 &0.0539 &0.1118 &0.2043 &0.2994 &0.3668 &0.1137 & \textbf{0.3855 } \\
    \bottomrule
    \end{tabular}
    }
    \label{table: region_task}
\end{table*}
\begin{table*}[ht!]
    \caption{Performance of GDHME on people-centric tasks.}
    \centering
    \resizebox{\columnwidth}{!} {
    \begin{tabular}{c|c|c|cccc|ccc}
    \toprule
    Task & Stage 2 & Metric & Random & Feature & Node2Vec & TGN & w EasyDec & w/o Static & GDHME \\
    \midrule
    \multirow{3}{*}{\textbf{Commuter}}& LR & AUC &0.4852 &0.6296 & OOM &0.5739 &0.5704 &0.5500 & \textbf{0.6336 } \\
    & KNN & AUC &0.4951 &0.5459 & OOM &0.5440 &0.5414 &0.5328 & \textbf{0.5523 } \\
    &MLP&AUC&0.4972 &0.5522 &OOM&0.5651 &0.5503 &0.5391 & \textbf{0.5672 } \\
    \hline
    \multirow{3}{*}{\textbf{Ride Hailing Driver}}& LR & AUC &0.5113 &0.7354 & OOM &0.5631 &0.5295 &0.4859 &\textbf{0.7644 } \\
    & KNN & AUC &0.5164 &0.5197 & OOM &0.5113 &0.4932 &0.4883 & \textbf{0.5364 } \\
    &MLP&AUC&0.4940 &0.6414 &OOM&0.5403 &0.5460 &0.4831 & \textbf{0.7059 } \\
    \hline
    \multirow{3}{*}{\textbf{House Owner}}& LR & AUC &0.5094 &0.6552 & OOM &0.5208 &0.5043 &0.5217 & \textbf{0.7698 } \\
    & KNN & AUC &0.4936 &0.5116 & OOM &0.4835 &0.5004 &0.5037 & \textbf{0.5850 } \\
    & MLP & AUC &0.4896 &0.5351 & OOM &0.5037 &0.5244 &0.5250 & \textbf{0.6658 } \\
    \hline
    \multirow{3}{*}{\textbf{Vehicle Owner}}& LR & AUC &0.5200 &0.5757 & OOM &0.5228 &0.4986 &0.4990 & \textbf{0.7001 } \\
    & KNN & AUC &0.5094 &0.5137 & OOM &0.5281 &0.5280 &0.4966 & \textbf{0.5619 }  \\
    & MLP & AUC &0.5004 &0.5142 & OOM &0.5538 &0.5311 &0.5094 & \textbf{0.5903 }  \\
    \hline
    \multirow{3}{*}{\textbf{APP Preference}}& LR & AUC &0.5071 &0.5211 & OOM &0.6103 &0.6068 &0.5517 & \textbf{0.6172 }  \\
    & KNN & AUC &0.4982 &0.5253 & OOM &0.5512 &0.5535 &0.5268 & \textbf{0.5755 }  \\
    & MLP & AUC &0.4959 &0.5138 & OOM &0.5701 &0.5557 &0.5157 & \textbf{0.5706 } \\
    \hline
    \multirow{4}{*}{\textbf{Traffic Preference}} & \multirow{2}{*}{KNN}& micro-F1 &0.1979 &0.2279 & OOM &0.2212 &0.2206 &0.2109 & \textbf{0.2416 } \\
    && macro-F1 &0.1931 &0.2218 & OOM &0.1709 &0.1725 &0.2115 & \textbf{0.2401 } \\
    \cline{2-10}
    & \multirow{2}{*}{MLP}& micro-F1 &0.1911 &0.2216 & OOM &0.2208 &0.2472 &0.2207 & \textbf{0.2665 } \\
    && macro-F1 &0.1897 &0.2118 &OOM&0.2201 &0.2461 &0.2201 &\textbf{0.2674 }\\
    \bottomrule
    \end{tabular}
    }
    \label{table: people_centric}
\end{table*}

\subsection{Comparison Results}
The \textbf{Random} method generates node representations as totally random vectors, which reflects the performance one can obtain without human mobility information.
The \textbf{Feature} method extract hand-designed features for each node.
For regions, we use the amount within an hour as one channel and features of all the time is a 240-dim vector.
For people, we calculate the amount of visited regions, the maximum visited time and the average visited time of one person as its spatial and temporal features.
\textbf{Node2Vec} is a static graph learning method.
\textbf{TGN} is a continuous-time dynamic graph learning method.
\textbf{w EasyDec} is GDHME trained with random negative samples.
\textbf{w/o Static} removes the module and only maintains dynamic memories for each node.

\textbf{Region-Centric Comparison}
The performance on region-centric tasks are shown in Table \ref{table: region_task}.
The first task about region is housing price prediction.
The data is collected online and the task reflects the comprehensive popularity of regions, including traffic convenience, administrative division and centrality.
The second task is shopping PoI amount prediction, which evaluates the ability to sense certain region attributes.
The third task is the prediction of the most PoI type in a region, which demands a comprehensive understanding of region functions.

It can be observed from the results that GDHME can achieve the best performance in all region-centric tasks, showing its effectiveness of learning region representations from human mobility data.
First, compared to Random and Feature, node representations from GDHME can capture more useful information with next location prediction pretext task.
The reason may be that this pretext task can extract more semantic from people-region interaction instead of merely macroscopic crowd flow trend.
Second, compared to Feature, learning-based methods can achieve better performance, indicating that though introducing human mobility in a naive way can be better than Random, it still needs a powerful method to discover complex patterns from data.
Third, designs for both harder tasks and static attributes in Stage 1 benefit the downstream tasks.
\textbf{People-Centric Comparison}
The results for people-centric tasks are shown in Table \ref{table: people_centric}.
These tasks are much harder than region-centric tasks in three reasons.
First, the people amount is much larger and their characteristics are much more diverse, which brings challenges to discriminate different types of people.
Second, the true labels of people are more expensive and harder to collect. Unlike region properties covering all the space, the people labels are always collected from a limited number of volunteers.
Third, compared to region nodes, individual behavior could be affected more by noises, anomaly data and happenchances.

From the results, we can observe that:
First, GDHME can achieve steady improvement in most tasks with limited labels.
It indicates that GDHME is capable of capturing individual semantics from human mobility with a task-agnostic pretext task and the ability benefits widely in applications where no extra people features can be obtained.
Second, similar to the regions, GDHME can benefit from both harder tasks and static attributes in most people-centric tasks.
Third, the Feature based method achieve a better performance in ride hailing driver classification. 
The reason is that this task highly relies on the basic features, such as if the people travel beyond a certain distance.
However, it perform less satisfactory on other semantic-demanding tasks.

\begin{figure}[!htbp]
    \centering
    \includegraphics[width=\columnwidth]{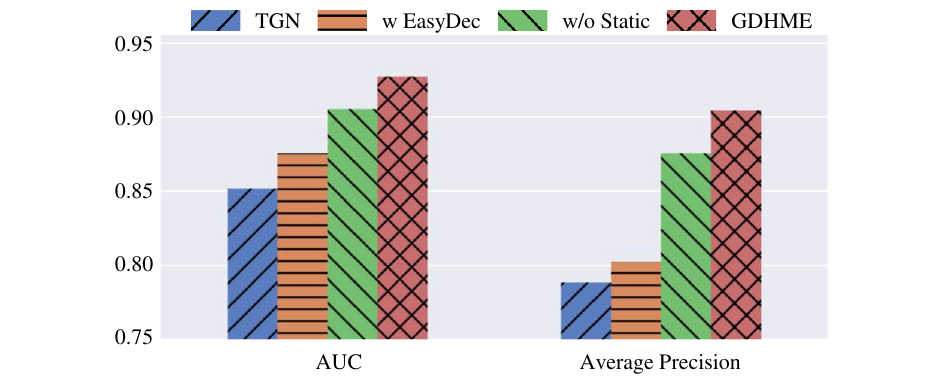}
    \caption{Performance on next location prediction.}
    \label{fig: ablation}
\end{figure}
\textbf{Next Location Prediction}
GDHME is also evaluated on next location prediction task.
The people-region interaction prediction can be helpful multiple applications including traffic management and personalized recommendation.
The task generates negative samples as described in Decoder part.
From Figure \ref{fig: ablation}, we can observe that both harder tasks and static attributes designs help GDHME predict next location more accurately.
\begin{figure*}[!htbp]
    \centering
    \includegraphics[width=\textwidth]{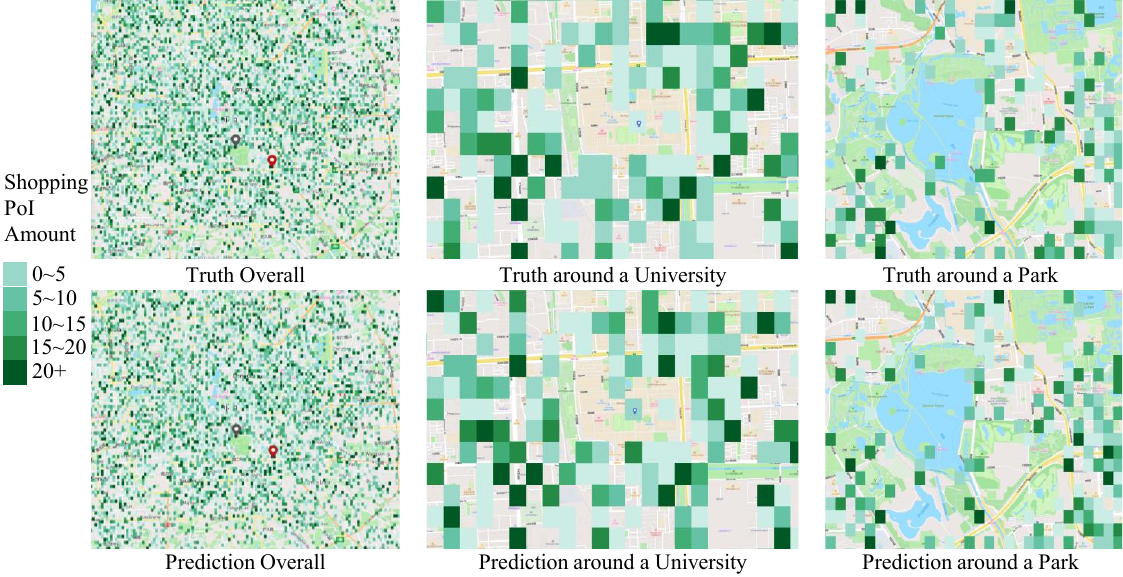}
    \caption{The illustration of shopping PoI amount prediction.}
    \label{fig: shopping_num}
\end{figure*}
\subsection{Illustration of Prediction}
To demonstrate the shopping PoI prediction intuitively, the prediction results are illustrated and shown in Figure \ref{fig: shopping_num}.
The three figures on the top row is drawn with the truth value of shopping point-of-interest amount and those on the bottom row is drawn with the prediction results.
The left column shows a global view and other two columns show local views.
The darker a grid is, the larger the value of the shopping point-of-interest is.
It can be observed that:
First, for the university area, the shopping points are sparse in the middle and denser around the university.
Second, for the park area, the shopping points are sparse around it but they are denser in the southwest and east.
Third, generally, the predicted amount corresponds to the truth value. In other words, where the shopping points in real world are dense, the prediction results with node representations are also larger.
The results shows that for different type of regions, the model can precisely learn the amount of shopping points, which indicates that the region node representations automatically extract spatial semantic information from only the human mobility data.
To some extent, this kind of node representations may be more useful than the hand-designed features.
For example, a comprehensive store may attract more people than two small stores.
When we analyse the region for shopping, the counting method may even mislead further applications.

\section{Deployment and Applications}

\subsection{Model Deployment}
GDHME has also been unveiled at the 2023 China Mobile Worldwide Partner Conference as the JiuTian ChuanLiu Big Model. 
The system, as illustrated in Figure \ref{fig: app_system}, has been deployed on the Jiutian Artificial Intelligence Platform, which was developed by China Mobile. 
This platform seamlessly integrates advanced artificial intelligence capabilities to empower businesses and individuals with intelligent, data-driven insights. 
Within this ecosystem, GDHME serves as a foundation model that extracts general-purpose features from massive-scale human mobility data to support a wide range of services, including business location selection, urban planning, and precision marketing. 

\begin{figure}[!htbp]
    \centering
    \includegraphics[width=0.9\columnwidth]{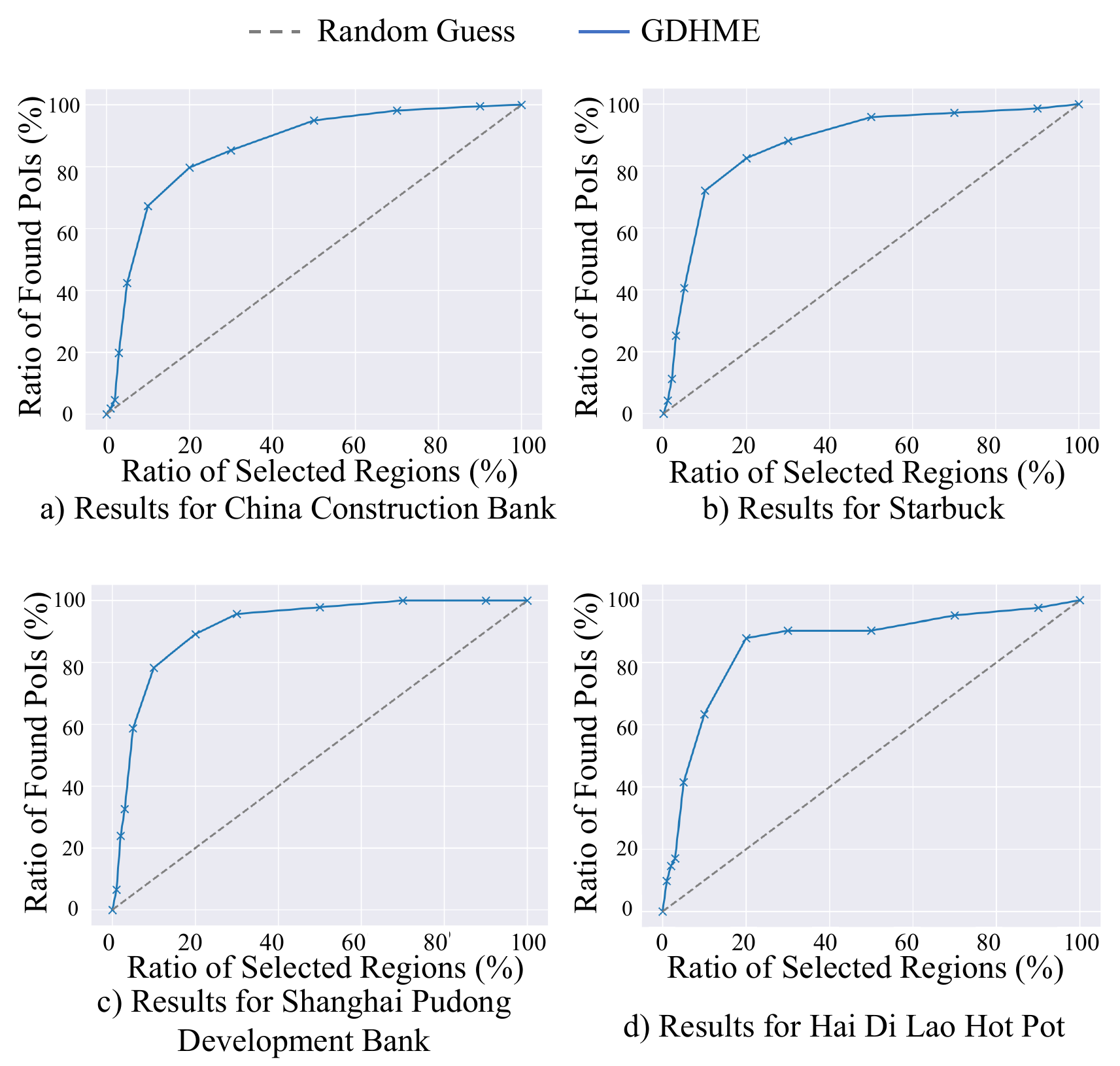}
    \caption{Example of GDHME-enabled applications for discovering points of interest (PoIs).}
    \label{fig: application}
\end{figure}

\subsection{Business-oriented Applications}
One representative application focuses on business location recommendation. 
In real-world scenarios, companies expanding their chain stores or seeking overlooked high-value locations often require substantial effort to investigate possible sites across an entire city. 
To address this need, we develop a service that accepts example store locations as input and generates candidate regions as potential new sites. 

For evaluation, we collect store locations from multiple chain businesses in a metropolitan area. 
We then use part of these locations as training samples for downstream models, such as Linear Regression, and predict whether the remaining regions are likely to contain valid store addresses. 
As illustrated in Figure \ref{fig: application}, the results show that a small ratio of recommended regions can cover a large fraction of actual store addresses. 
We further validate this on different businesses, including two banks and two chain restaurants, all demonstrating consistent effectiveness. 

Through GDHME, users can access the system and immediately obtain valuable location recommendations by simply inputting a few reference locations. 
The framework automatically discovers latent connections between candidate and reference sites, driven by in-depth human mobility patterns captured in the learned representations. 

\subsection{People-centric Applications}
Another application scenario addresses people-centric analysis, which is highly relevant for industries such as insurance and mobility services. 
Powered by the learned people representations from GDHME, we can construct labeled datasets and train predictive models for various user attributes. 
For instance, as shown in Table \ref{table: people_centric}, GDHME representations are used to identify individuals who are more likely to own a house or serve as ride-hailing drivers. 
This capability enables companies to conduct targeted marketing campaigns and service allocation with much higher efficiency. 

\textbf{Notably, the entire procedure can be completed within minutes using GDHME, whereas traditional methods may require dozens of personnel and several months of investigation. In addition, GDHME provides a reliable foundation based on large-scale human mobility data, ensuring both accuracy and scalability.}

\subsection{Illustration of the System}
\begin{figure}[!htbp]
    \centering
    \includegraphics[width=\columnwidth]{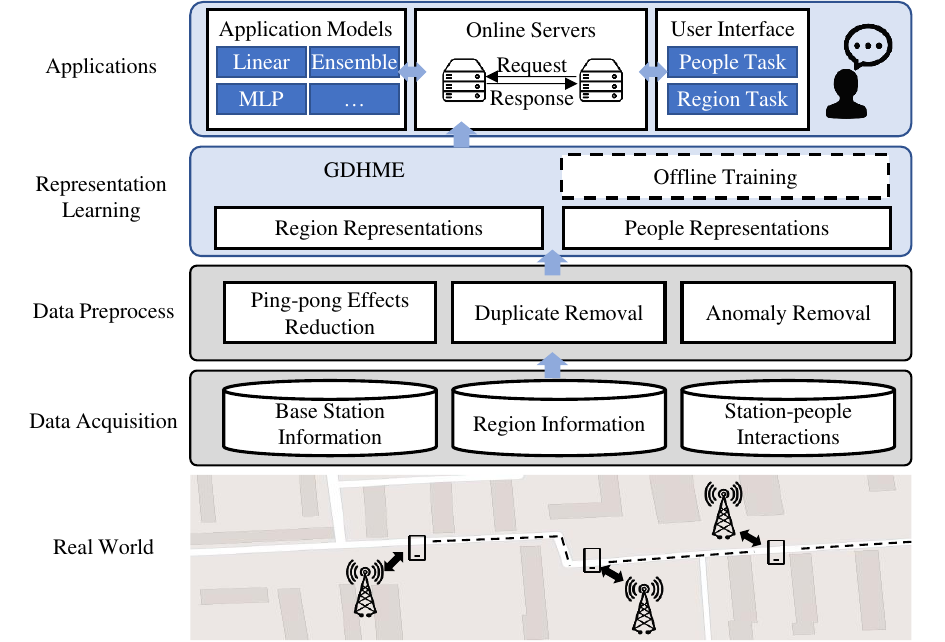}
    \caption{The overview of JiuTian·Chuanliu.}
    \label{fig: app_system}
\end{figure}
\begin{figure}[!htbp]
    \centering
    \includegraphics[width=\columnwidth]{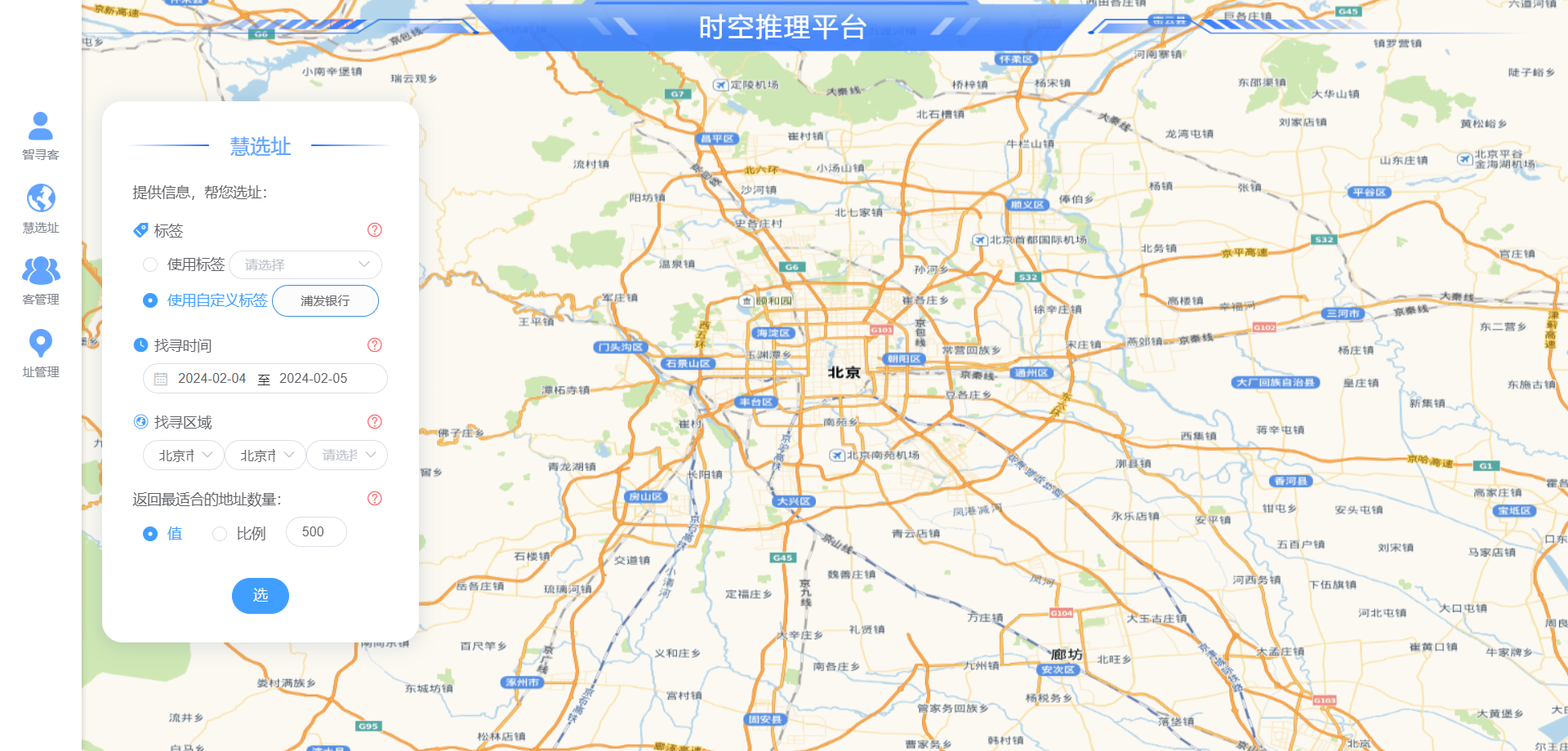}
    \caption{The user input interface.}
    \label{fig: app_userinput}
\end{figure}
\begin{figure}[!htbp]
    \centering
    \includegraphics[width=\columnwidth]{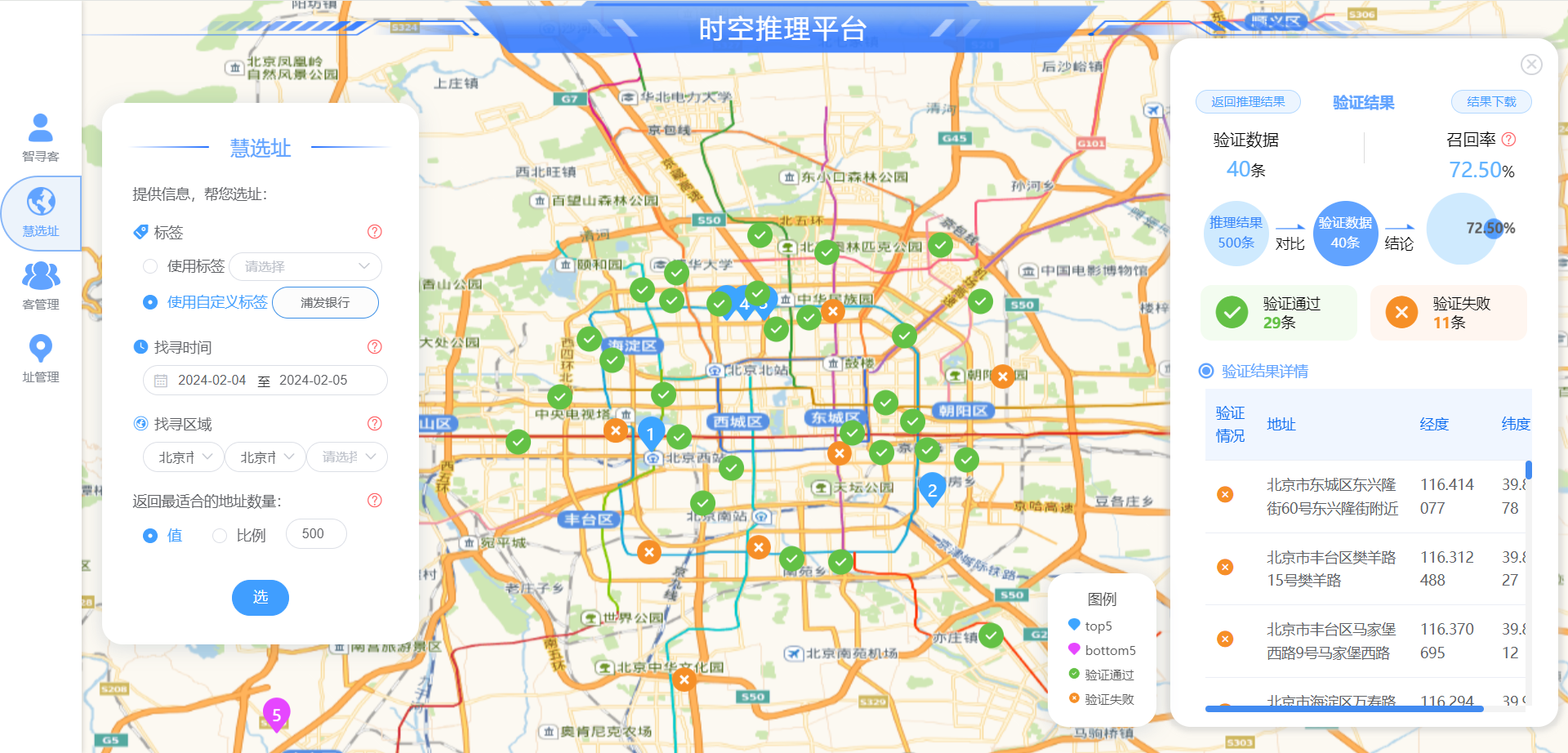}
    \caption{The results of location selection for banks.}
    \label{fig: app_bankoutput}
\end{figure}
\begin{figure}[!htbp]
    \centering
    \includegraphics[width=\columnwidth]{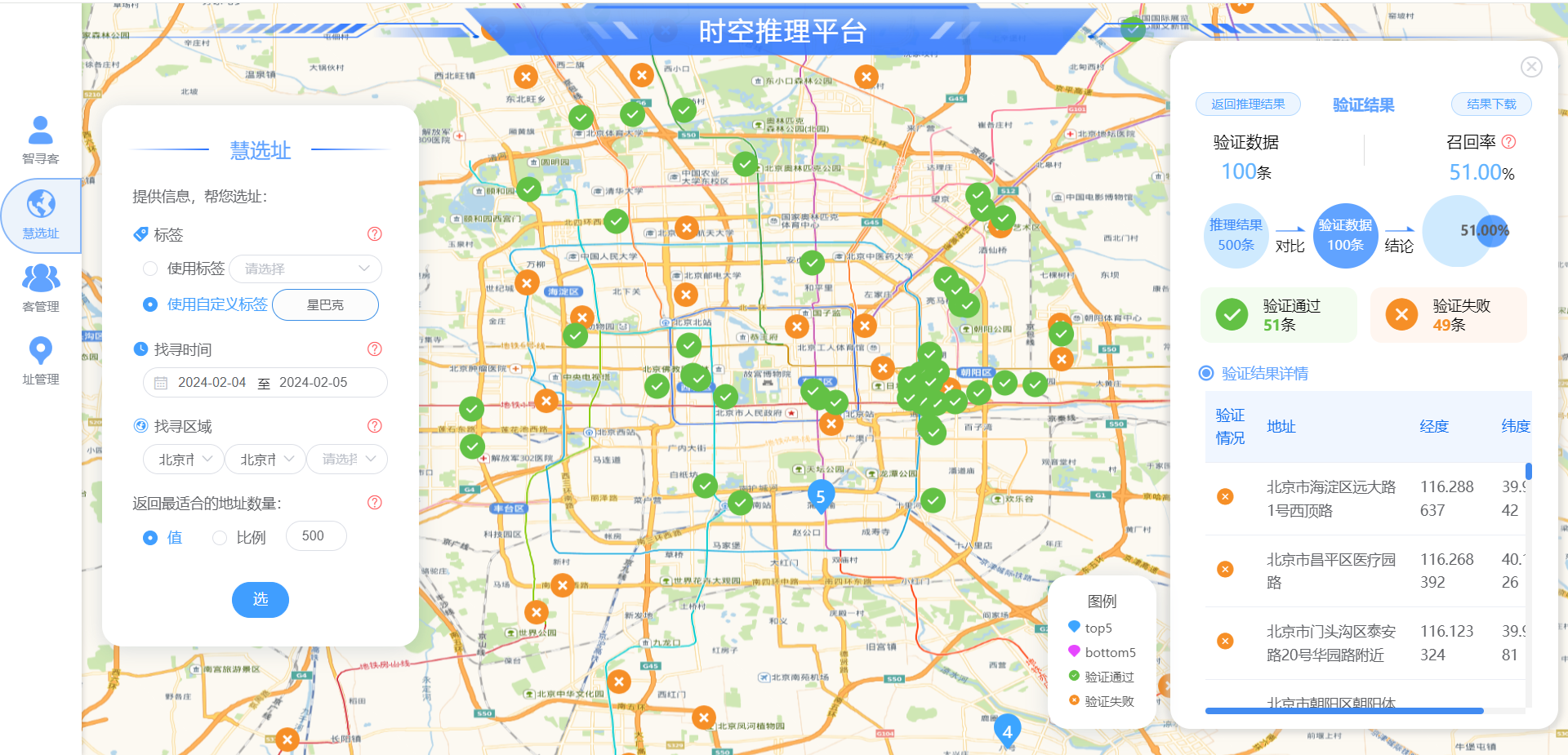}
    \caption{The results of location selection for cafes.}
    \label{fig: app_starbuck}
\end{figure}

As shown in Figure \ref{fig: app_system}, a system powered by GDHME is also developed and deployed.
The user interface is illustrated in Figure \ref{fig: app_userinput}, Figure \ref{fig: app_bankoutput} and Figure \ref{fig: app_starbuck}.
With the above applications as the core, this system can receive personalized input from the user, which implicitly describes the task, and generate task-specific results.
First, the user can collect some training labels about the target task.
For example, if the target task is to select locations for a new bank, then the training labels can be the locations of currently opening banks.
Then these data are input through an interface shown in Figure \ref{fig: app_userinput}.
There data can tell the model which task to solve.
The procedure is also similar for regression tasks, binary-classification tasks and multi-classification tasks.
Second, with the input labels, the system will obtain well trained node representations as node features and train a downstream model to fit the target task.
Then, the model can generate labels for all other regions or people.
The rationale is that the node features already contains the information from large-scale human mobility data.
If some examples are given, we can find regions or people having similar features with these examples and make predictions for specific tasks.
Finally, the results are returned to the user.
Shown in Figure \ref{fig: app_bankoutput} and Figure \ref{fig: app_starbuck} are results for bank location selection and cafe location selection respectively.

From the results, it can be observed that for different tasks, the model can automatically generate adaptive results.
For the bank, the locations tend to be distributed all across the city; and for the cafe, the locations tend to gather around the working areas.
Correspondingly, it can also be observed that the locations selected by the model can also cover a large portion of locations that contains an actual bank or cafe.
Compared with the traditional workflow, the GDHME can save a lot of manpower and introduce new knowledge from massive human mobility data.

\begin{figure}[!htbp]
    \centering
    \includegraphics[width=\columnwidth]{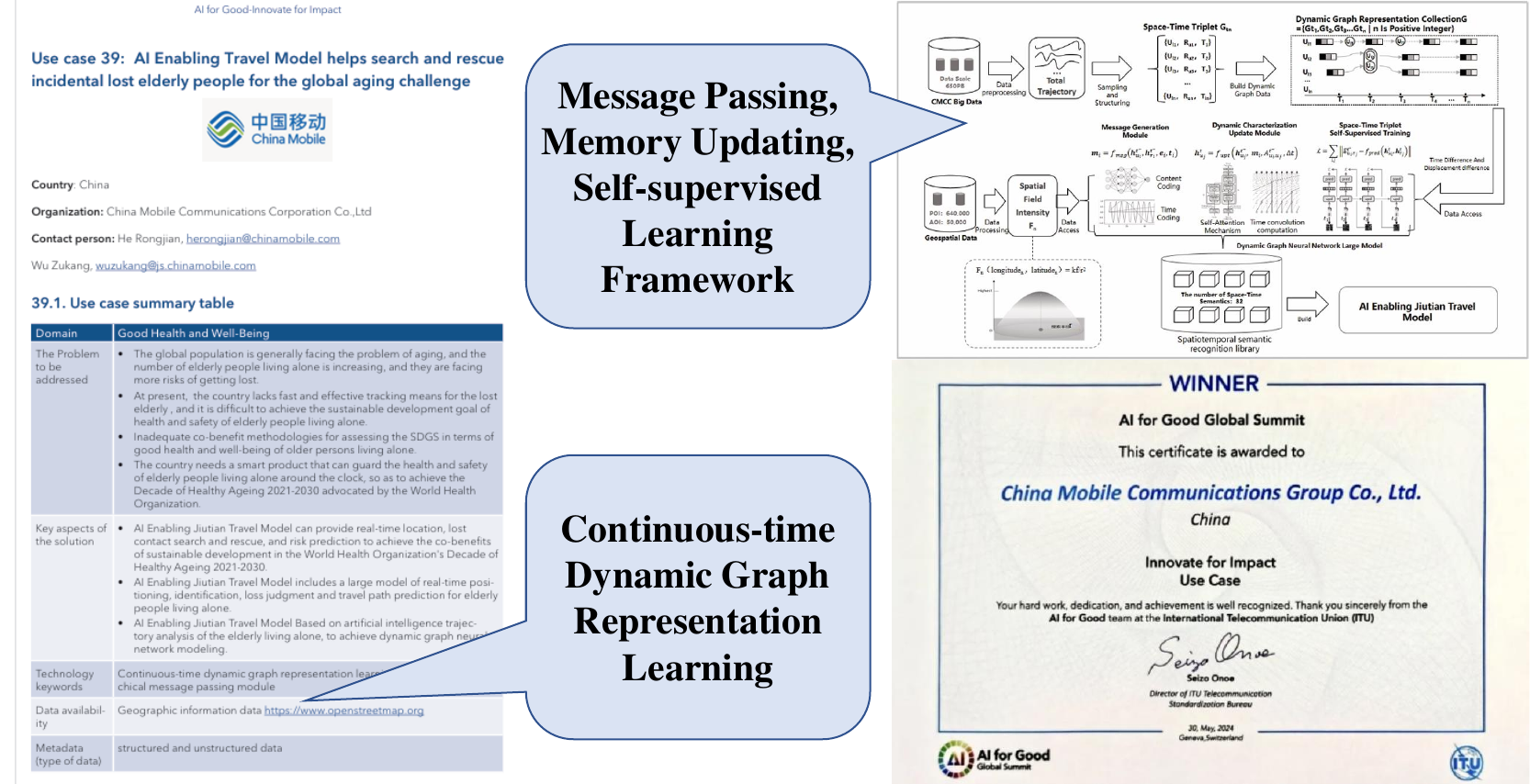}
    \caption{The Innovate for Impact User Case Award from the AI for Good Summit.}
    \label{fig: app_award}
\end{figure}

Moreover, applications based on continuous-time dynamic graph representation learning with the human mobility data also won the Innovate for Impact User Case Award from the AI for Good Summit as shown in Figure \ref{fig: app_award}\footnote{Use Case 39 at \url{https://s41721.pcdn.co/wp-content/uploads/2021/06/2400805_Use-cases-collection.pdf}}.

\section{Related Work}
\subsection{Human Mobility Modeling}
In recent years, researchers have proposed various methods to learn from human mobility.
From region view, some recent studies \cite{DBLP:conf/ijcai/WuYFPZZ0W22,wang2017region,DBLP:conf/ijcai/0004LLH20,DBLP:conf/www/ZhangHXWLY23,DBLP:conf/aaai/ZhouHCS023} combine human mobility-based graphs with graphs from multiple types of data, to generate static region embeddings.
Some researchers compute crowd flow from raw human mobility and learn region states for future flow prediction.
Early studies \cite{yao2018deep, Deepresnet, revisiting, ye2019co} divide the studied areas to learn the states of grid.
Later, several methods \cite{stgcn,dcrnn,gwnet,mtgnn,agcrn,astgcn,DBLP:conf/kdd/HanDSFL021,DBLP:conf/aaai/YeSDF021} powered by graph neural network are proposed for the situation where the regions are arbitrary shaped.
These methods aim at solving region-level tasks, omitting the fine-grained semantic of the people.
From people view, most researchers extracts trajectories from original data and learn the behavior feature of people.
Generally, most researchers encode the trajectories by either geographical locations \cite{DBLP:conf/ijcai/Zhang0SZ18,DBLP:conf/aaai/WangZCLZ18} or topology locations \cite{DBLP:conf/kdd/WangFY18,DBLP:conf/aaai/LinW0L21,DBLP:journals/tkde/ZhaoLLXLZSZ22}.
These methods aim at solving people-level tasks, while they always omit the dynamic states of the regions and encode them by relatively simple way.

\subsection{Graph Representation Learning}
Traditionally, graph representation learning focused on static graphs where nodes and edges are fixed \cite{GCN, GraphSage, GAT, GIN}.
These methods would fail for many scenarios where the graph would evolve.
Recent years have witnessed a branch of studies for dynamic graphs.
These studies they can be roughly divided into two categories: discrete-time dynamic graphs (DTDGs) and continuous-time dynamic graphs (CTDGs).
Methods for DTDGs views dynamic graphs as a combination of multiple static graphs; each static graph represents topology information in a time window \cite{DynGEM, cc, DySAT, EvolveGCN}.
These studies take the evolving factors into consideration, while they still loss some fine-grined time information and are still inflexible in some cases.
Methods on CTDGs views the basic elements of dynamic graphs as timestamped edges \cite{CTDNE, DyRep, JODIE, STREAMING, TGAT, tgn, TagGen}.
These edges indicates the interaction between nodes and the basic idea is to compress the historical interaction information into involving nodes.
Although they can handle the interactions between nodes, how it performs in human mobility modeling remains unexplored.
\section{Conclusion}
In this paper, we present a evolving graph representation learning framework, called GDHME, which can automatically extract abundant semantics from large-scale human mobility data.
GDHME first unifies human mobility data as people-region interactions to keep the original information as much as possible.
Then, we design a continuous-time dynamic encoder to compress historical interaction records into node memories.
Moreover, considering the unique problem in human mobility, an enhanced decoder is specially designed.
Extensive experiments are conducted on a large-scale real-world signaling dataset, which has a wide coverage in both regions and people.
This study also constructs a multi-task benchmark to evaluate the general-purpose region and people representations.
An online service JiuTian ChuanLiu is also deployed to support real demand.
This pioneering study could provide new insights for diverse smart city applications.

\textbf{Limitations and Future Work}. Despite the promising results, several limitations should be acknowledged. First, the learned representations are derived solely from spatiotemporal mobility behaviors. While effective for many urban sensing tasks, they may not fully capture the specific attributes required in certain domains, such as fraud detection, where communication logs or behavioral records are indispensable. Second, the framework may face challenges in modeling rare or highly irregular mobility patterns, as these cases are often underrepresented in large-scale datasets. These limitations suggest directions for future research, including the integration of heterogeneous data sources (e.g., communication records, socio-economic indicators, or environmental data) to enrich the semantic space and enhance the generality of the learned embeddings.
\bibliographystyle{ACM-Reference-Format}
\bibliography{sample-base}

\end{document}